\documentclass[doublecol,linenumbers]{epl2}
\usepackage{graphicx,bm,amsmath,amssymb}
\DeclareMathOperator{\Tr}{Tr}
\newcommand{\up}{\uparrow}
\newcommand{\dn}{\downarrow}

\title{Crossover from collisionless to collisional spin dynamics of polarized fermions}
\shorttitle{crossover from collisionless to collisional spin dynamics of polarized fermions}

\author{Junjun Xu \and Qiang Gu \thanks{Email: \email{qgu@ustb.edu.cn}}}
\shortauthor{Junjun Xu \etal}

\institute{Department of Physics, University of Science and Technology Beijing, Beijing 100083, China}


\abstract{We study the transverse spin dynamics of trapped polarized Fermi gases in the high temperature limit. In the non-interacting collisionless regime, a magnetic field gradient induces collective spin wave oscillations. In the strongly interacting collisional regime, the dynamics are governed by spin diffusion. These two limits have been extensively studied both experimentally and theoretically, but the crossover between them has received less attention. In this paper, we use a quantum Boltzmann equation to study transverse spin dynamics and show how the excitations evolve from dispersive to diffusive in the high temperature limit. We provide analytical solutions in the two limiting regimes, which agree well with our numerical results.}

\begin{document}

\maketitle

\section{Introduction}
Recently the transverse spin dynamics of cold fermionic atoms has attracted much attention. In the non-interacting collisionless regime, a magnetic field gradient induces collective spin wave oscillations. In the strongly interacting collisional regime, the dynamics are governed by spin diffusion. These two regimes have been studied experimentally and theoretically in cold atoms \cite{Kohl, Thywissen1, Thywissen2, Enss1, Enss2, Xu, Koller, Sommer1, Sommer2, Levin, Enss3, Bruun1, Bruun2, Hild, Koller2}, liquid helium \cite{Leggett, Lhuillier, Mullin1, Mullin2}, and solid state systems \cite{Solid1,Solid2,Solid3}.

Despite the extensive studies of these two regimes, one natural question remains: how does spin transport occur as the system evolves from collisionless to collisional? There have been very few studies of the crossover between these regimes. Cold atoms are ideal for exploring this physics, as one can continuously change the interaction strength \cite{Ketterle, Bloch, Chin}. Here we address the collisionless to collisional crossover in the high temperature limit.

In the following, we consider a two-component Fermi gas placed in a cigar shape trap. The long axis is denoted $\hat{z}$. Following the protocol in the Toronto experiments \cite{Thywissen1}, we model a gas which is initially prepared with a uniform magnetization in the $\hat{y}$ direction. A magnetic field is applied in the $\hat{z}$ direction. The strength of this field varies linearly with $z$. The magnetic field gradient causes the spins to form a helix. Spin waves and spin diffusion influences these dynamics. One probes their role through a spin-echo technique. One applies a $\pi$-pulse at time $t=t_\pi$, flipping each spin so that $S_x\to-S_x$, $S_y\to-S_y$, $S_z\to-S_z$. This reverses the helicity of the spin texture. At time $t=2t_\pi$ the total magnetization is measured. In the absence of transport, the final magnetization is equal to the initial magnetization. As such, any reduction in the final magnetization gives information about transport. Since most of the relevant dynamics are one dimensional (1D), we will focus on a 1D model.

This paper is organized as follows. First we present our model and analyse it in the absence of interactions. We find how the collisionless spin waves influence the experiment. Then in the collisional regime, we solve a Fokker-Planck-like diffusion equation, finding analytical results for the influence of spin diffusion on this experiment. At last, we show how the system evolves form collisionless to collisional by numerically solving the linearised Boltzmann equation.

\section{The model and collisionless regime: spin wave oscillations}
We consider 1D fermions confined in a harmonic trap. The system is initially polarized in the $\hat{y}$ direction and then a magnetic field gradient is added in the $\hat{z}$ direction to rotate the spins. The Hamiltonian of the system is
\begin{align}
H=&\sum_{\sigma=\uparrow,\downarrow}\int\psi_\sigma^\dagger(x)
\left(-\frac{\partial_x^2}{2}+\frac{x^2}{2}+\frac{\lambda}{2}\gamma_\sigma x\right)\psi_\sigma(x)dx\nonumber\\
&+g\int\psi_\uparrow^\dagger(x)\psi_\downarrow^\dagger(x)\psi_\downarrow(x)\psi_\uparrow(x)dx,
\end{align}
where $\gamma_\sigma=\mp1$ for the $\up$ and $\dn$ spins. We have used the dimensionless quantities $x=\tilde{x}/l$, $\lambda=\tilde{\lambda}l/(\hbar\tilde{\omega})$, $g=\tilde{g}/(\hbar\tilde{\omega} l)$. Here the physical quantities are labelled with a tilde and $l=\sqrt{\hbar/(m\tilde{\omega})}$ is the characteristic length associated with the harmonic trap of frequency $\tilde{\omega}$. $\tilde{\lambda}=\Delta\mu\partial_xB$ characterizes the Zeeman energy induced by the magnetic field gradient with $\Delta\mu$ the corresponding magnetic moment.

The dynamics of the system are encoded in the semi-classical distribution function
\begin{align}
f_{\sigma\sigma'}(x,v,t)=\frac{1}{2\pi}\int\rho_{\sigma\sigma'}\left(x+\frac{r}{2},x-\frac{r}{2},t\right)e^{ivr}dr,
\label{eq:f}
\end{align}
where $t=\tilde{\omega}\tilde{t}$ is the dimensionless time, $x$ is the position, $v$ is the velocity, and the correlation function $\rho_{\sigma\sigma'}(x_1,x_2,t)=\left<\psi_\sigma^\dagger(x_1,t)\psi_{\sigma'}(x_2,t)\right>$, with $\left<...\right>$ the thermal ensemble. Physically, $f_{\sigma\sigma'}$ encodes the distribution of particles in phase space. The correlation function evolves according to the equation of motion: $\dot{\rho}_{\sigma\sigma'}(x_1,x_2,t)=i\left<[H,\psi_\sigma^\dagger(x_1,t)\psi_{\sigma'}(x_2,t)]\right>$, which will give us the Boltzmann equation for $f_{\sigma\sigma'}$. In particular, the transverse distribution function $f_\perp=f_{\up\dn}$ obeys
\begin{align}
(\partial_t+v\partial_x-x\partial_v+i\lambda x)f_\perp(x,v,t)=C_\perp[\vect{f}].
\label{eq:Boltz}
\end{align}
The second term on the left accounts for drift, and is the origin of spin diffusion. The third and forth terms on the left account for forces from the trap potential and magnetic field. The transverse collision integral $C_\perp[\vect{f}]=C_{\up\dn}[\vect{f}]$ describes the change of phase space density due to collision. Here $\vect{f}$ are $2\times2$ matrices with elements $f_{\sigma\sigma'}$.

We first consider the non-interaction limit, where $C_\perp[\vect{f}]=0$. Before adding the gradient magnetic field, the system is in equilibrium with all the spins polarized in the $\hat{y}$ direction. We consider a high temperature limit with $f_\perp(x,v,0)=e^{(\mu-x^2/2-v^2/2)/T}$, where $\mu=\tilde{\mu}/(\hbar\tilde{\omega})$ and $T=k_B\tilde{T}/(\hbar\tilde{\omega})$ are dimensionless chemical potential and temperature. The spin echo technique is implemented by using a $\pi$ pulse at half of the evolution time, i.e., flipping the direction of the gradient magnetic field $\lambda\to -\lambda$ at $t=t_\pi$ and then measuring the magnetization at time $t=2t_\pi$. We find that the normalized transverse magnetization is (see the appendix)
\begin{align}
M_e(t=2t_\pi)=\frac{1}{2\pi N}\int f_\perp(x,v,t)dxdv=e^{-8\lambda^2T\sin(t/4)^4},
\label{eq:nonint}
\end{align}
where $N=Te^{\mu/T}$ is the particle number. This formula agrees with the high temperature limit of the result in \cite{Koller}.

\begin{figure}[h]
  \centering
  \includegraphics[width=0.4\textwidth]{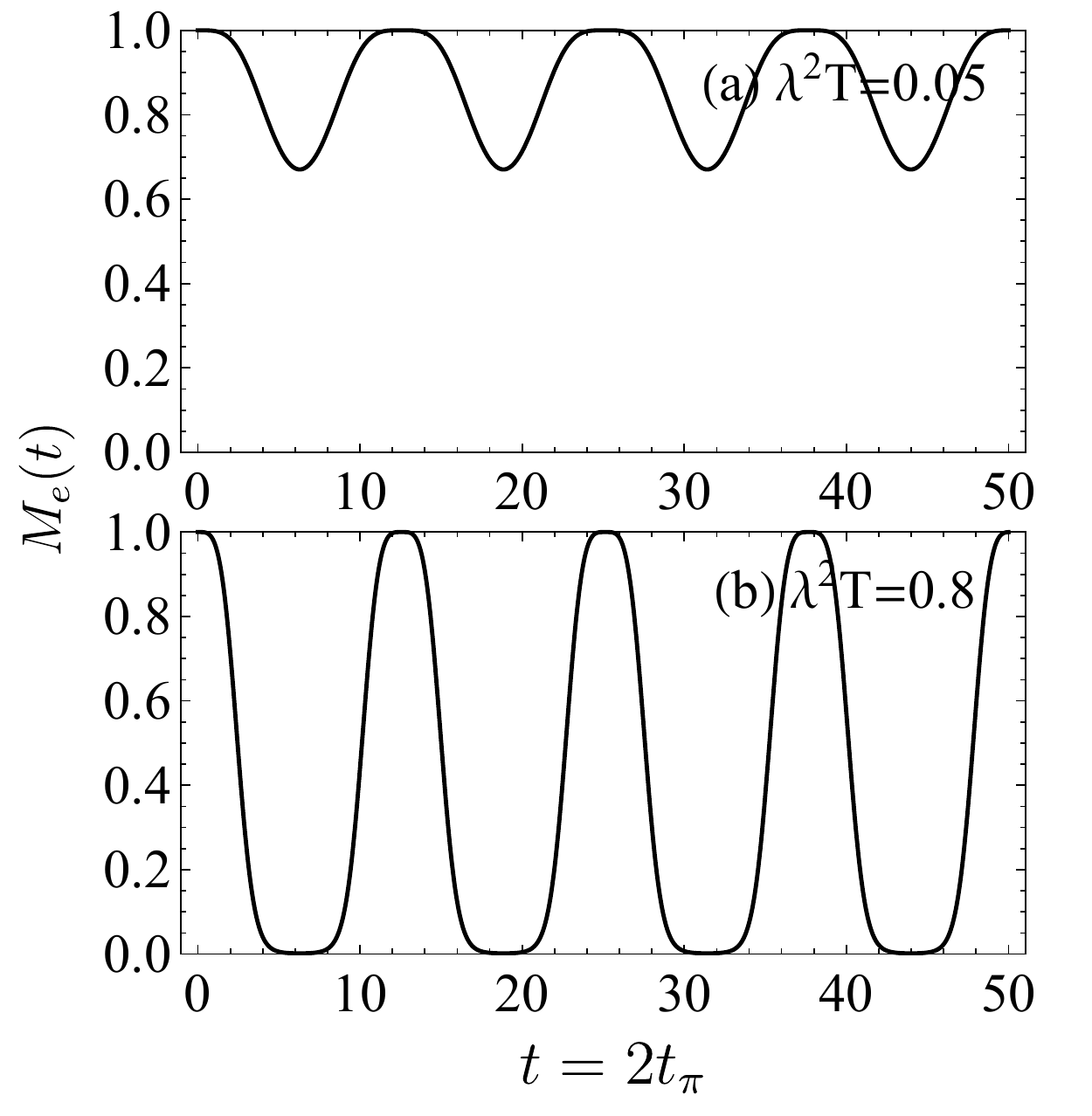}\\
  \caption{Transverse magnetization $M_e(t=2t_\pi)$ evolution as a function of time $t=\tilde{\omega}\tilde{t}$ in a non-interacting collisionless limit. Here we choose $\lambda^2T=k_B\tilde{T}\tilde{\lambda}^2l^2/(\hbar\tilde{\omega})^3=0.05$ (a) and $0.8$ (b).}\label{fig:fig1}
\end{figure}

This magnetization is plotted in Fig. \ref{fig:fig1} for two values of $\lambda^2T$. It displays oscillations which are similar to those in \cite{Xu, Koller}. The oscillation amplitude increases for larger magnetic field gradient and the magnetization reaches its minimum $\exp(-8\lambda^2T)$ at time $t=\tilde{\omega}\tilde{t}=2(2n+1)\pi$, where $n\in\mathbb{N}$. Here we see the normalized magnetization is independent of particle number, and hence the dynamics are independent of system size. This behavior is very different from the zero temperature case, where the oscillation amplitude increases as the particle number increases \cite{Xu}. As we show in the following sections, interactions cause these oscillations to decay. This decay is the signature of spin diffusion.

\section{The collisional regime: spin diffusive dynamics}
Spin diffusion dominates in the collisional limit. In a relaxation time approximation, the transverse collision integral is $C_\perp[f_\perp]=-f_\perp(x,v,t)/\tau$, where $\tau=\tilde{\omega}\tilde{\tau}$ is the reduced scattering lifetime with $\tilde{\tau}$ the physical one. By calculating the zeroth and first moment of Eq. (\ref{eq:Boltz}), one finds
\begin{align}
&\partial_tm+\partial_xJ+i\lambda xm=0,\\
&\partial_tJ+\bar{v}^2\partial_xm+xm+i\lambda xJ=-J/\tau,
\label{eq:momentum}
\end{align}
where $m(x,t)=1/(2\pi)\int f_\perp(x,v,t)dv$ is the transverse magnetization density and $J(x,t)=1/(2\pi)\int vf_\perp(x,v,t)dv$ is the corresponding fluid density. We have used $\int v^2f_\perp(x,v,t)dv\approx\bar{v}^2\int f_\perp(x,v,t)dv$, where $\bar{v}^2$ describes the mean kinetic energy of the system. When $\tau$ is small, the current $J$ is also small, and one can neglect the terms $\partial_tJ$ and $i\lambda xJ$ in Eq. (\ref{eq:momentum}). This yields a Fokker-Planck-like diffusion equation\begin{align}
(\partial_t-D\partial_x^2+i\lambda x)m-\tau\partial_x(x m)=0,
\label{eq:trapdiff}
\end{align}
where the last term on the left accounts for the force by the harmonic trap and $D=\bar{v}^2\tau$ is the diffusion constant. We now solve Eq. (\ref{eq:trapdiff}) and calculate the resulting magnetization. In a high temperature limit, the system with distribution function $f_\perp(x,v,0)=e^{(\mu-x^2/2-v^2/2)/T}$ will have initial magnetization density $m(x_0,0)=Ne^{-x_0^2/(2T)}/\sqrt{2\pi T}$. If the same spin echo technique is applied at time $t_\pi$, we get the normalized transverse magnetization at $t=2t_\pi$ as (see the appendix)
\begin{align}
M_e(t=2t_\pi)=e^{-Q\lambda^2},\label{eq:solution}
\end{align}
where
\begin{align}
Q=\frac{e^{-2\tau t}}{2\tau^3}\left\{(-1+e^{\tau t/2})^4\tau T
+D[-1+4e^{\tau t/2}\right.\nonumber\\
\left.-8e^{\tau t}+12e^{3\tau t/2}+e^{2\tau t}(-7+2\tau t)]\right\}.
\end{align}
This result has not previously appeared in the literature.
\begin{figure}[h]
  \centering
  \includegraphics[width=0.48\textwidth]{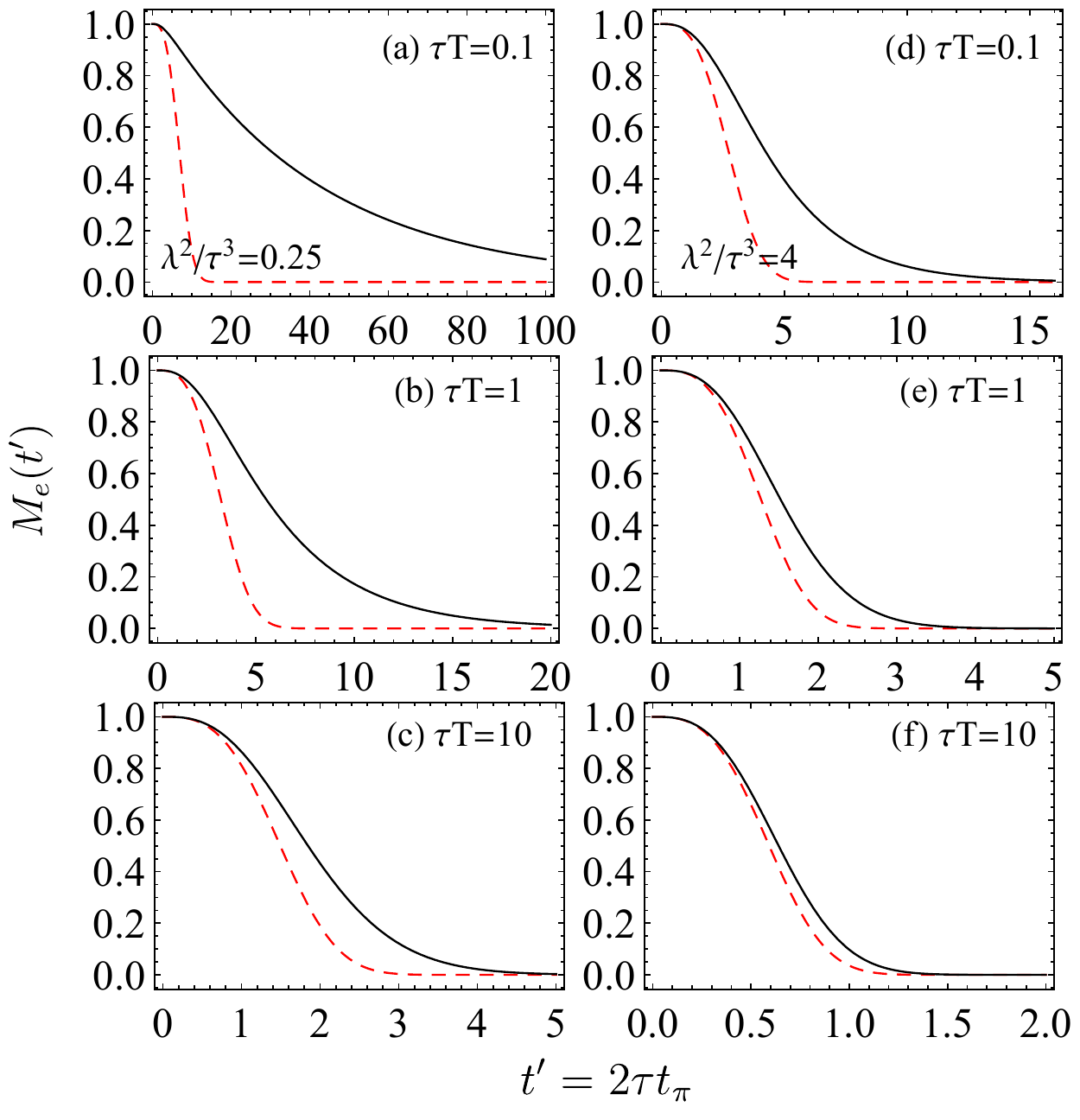}\\
  \caption{Transverse demagnetization in a harmonic trap for magnetic field gradient $\lambda^2/\tau^3=0.25$ from (a-c) and $4$ from (d-f) for temperatures $\tau T=0.1, 1,10$. The diffusion constant is $D=\tau T$. The dashed red lines are the homogeneous solutions $M_e^0(t=2t_\pi)=\exp(-D\lambda^2t^3/12)$ and the solid black lines correspond to our solution Eq. (\ref{eq:solution}).}
  \label{fig:fig2}
\end{figure}

In the limit of zero scattering lifetime $\tau\to 0$, we recover the homogeneous result $M_e(t)=\exp(-D\lambda^2t^3/12)$ \cite{Thywissen1}. For small $\tau$, we have $M_e(t)\approx\exp(-D\lambda^2t^3/12+D\lambda^2t^4\tau/16)$. In a quasi-equilibrium state, we have $D\approx\tau T$ and the Eq. (\ref{eq:solution}) is only a function of $t'=\tau t$, $T'=\tau T$, and $\lambda'^2=\lambda^2/\tau^3$. In Fig. \ref{fig:fig2} we show the magnetization for two values of the magnetic field gradient, $\lambda'^2=0.25$ in (a-c) and $\lambda'=4$ in (d-f), for three values of the temperature $T'=0.1, 1,10$. The deviation from the homogeneous solution (dashed red lines) becomes larger for higher magnetic field gradient $\lambda$ and lower temperature $T$ for fixed $\tau$.

\section{Crossover from collisionless to collisional regime}
To study the crossover between these two regimes, we solve the quantum Boltzmann equation beyond the relaxation time approximation. Mullin {\it et al}. derived the relevant collision integral \cite{Mullin1, Mullin2}. In Born approximation, it is written as \cite{Enss1}
\begin{align}
\vect{C}[\vect{f}_1]=&\frac{g^2}{8\pi}\int dk_2dk_3dk_4\delta(k_1+k_2-k_3-k_4)\nonumber\\
&\times \delta(k_1^2+k_2^2-k_3^2-k_4^2)\{[\bar{\vect{f}_1},\bar{\vect{f}}_2^-]_+\Tr(\vect{f}_3\vect{f}_4^-)\nonumber\\
& -[\vect{f}_1,\vect{f}_2^-]_+\Tr(\bar{\vect{f}_3}\bar{\vect{f}}_4^-)\}.
\end{align}
It describes processes where two fermions with momenta $k_1$ and $k_2$ scatter into momenta $k_3$ and $k_4$. Here $\vect{f}_m=\vect{f}(x,k_m,t)$ is the $2\times2$ matrix defined in Eq. (\ref{eq:f}), and $\bar{\vect{f}}=\vect{I}-\vect{f}$, $\vect{f}^-=\Tr(\vect{f})\vect{I}-\vect{f}$. The $[,]_+$ corresponds to the anticommutator.

During the dynamics the distribution function remains near
\begin{align}
f_{\sigma\sigma'}=n_{\sigma\sigma'}+\delta f_{\sigma\sigma'}
\end{align}
where $n_{\sigma\sigma'}$ is the equilibrium distribution function. Like previous section, we only focus on the high temperature limit, and we start with a initial distribution $f_{\up\up}=f_{\dn\dn}=f_{\up\dn}=f_{\dn\up}=e^{(\mu-x^2/2-v^2/2)/T}$. The collision between different spins reduces the total magnetization until the system reaches equilibrium with no magnetization. The equilibrium state distribution function is then written as $n_\up=n_{\uparrow\uparrow}=e^{(\mu_0-x^2/2-\lambda x/2-v^2/2)/T}$, $n_\dn=n_{\downarrow\downarrow}=e^{(\mu_0-x^2/2+\lambda x/2-v^2/2)/T}$, and $n_{\uparrow\downarrow}=n_{\downarrow\uparrow}=0$. So approximately we have
\begin{align}
\vect{f}\approx\left[
\begin{array}{cc}
n_\up & \delta f_{\up\dn} \\
\delta f_{\dn\up} & n_\dn \\
\end{array}
\right]
=\left[
\begin{array}{cc}
n_\up & f_\perp \\
f_\perp^* & n_\dn \\
\end{array}
\right].
\end{align}
To connect with the experiment, like the 1D experiment carried out by \cite{Hulet}, we choose particle number $N=Te^{\mu/T}\approx100$ with the temperature $T=\tilde{T}/(\hbar\omega)=500$ and $\mu=\tilde{\mu}/(\hbar\omega)=-800$. For a axial trap frequency $\omega=2\pi\times 200{\rm Hz}$ \cite{Hulet}, this corresponds to temperature $\tilde{T}\approx50\times10^3{\rm nk}$. Since the Fermi temperature is around $T_F\approx100\hbar\omega/k_B\approx 10\times 10^3{\rm nk}=\tilde{T}/5$, we are in a high temperature regime. As with the previous sections, one finds similar results for other parameters. We self-consistently find $\mu_0$ to fix $N=100$.

We consider a 1D system here, if we take into account only the on-shell contribution, we have $k_1=k_3,k_2=k_4$. Thus the dynamics of the transverse and longitudinal distribution functions are decoupled and we have the following Boltzmann equation
\begin{align}
(\partial_t+v\partial_x-x\partial_v+i\lambda x)f_\perp=C_\perp[f_\perp(v=k_1)],
\label{eq:linear}
\end{align}
with the linearised transverse collision integral
\begin{align}
C_\perp=&-\frac{g^2}{8\pi}\int dk_2\left(f_{1\perp}n_2-n_1f_{2\perp}\right)(\bar{n}_{2\up}\bar{n}_{1\dn}+\bar{n}_{2\dn}\bar{n}_{1\up})\nonumber\\
&-\left[f_{1\perp}(2-n_2)-(2-n_1)f_{2\perp}\right]\nonumber\\
&\times(n_{2\up}n_{1\dn}+n_{2\dn}n_{1\up}),
\end{align}
where $n_m=n(x,k_m,t)$, $f_m=f(x,k_m,t)$, and $n=n_{\dn}+n_{\up}$.

\begin{figure}[h]
  \centering
  \includegraphics[width=0.48\textwidth]{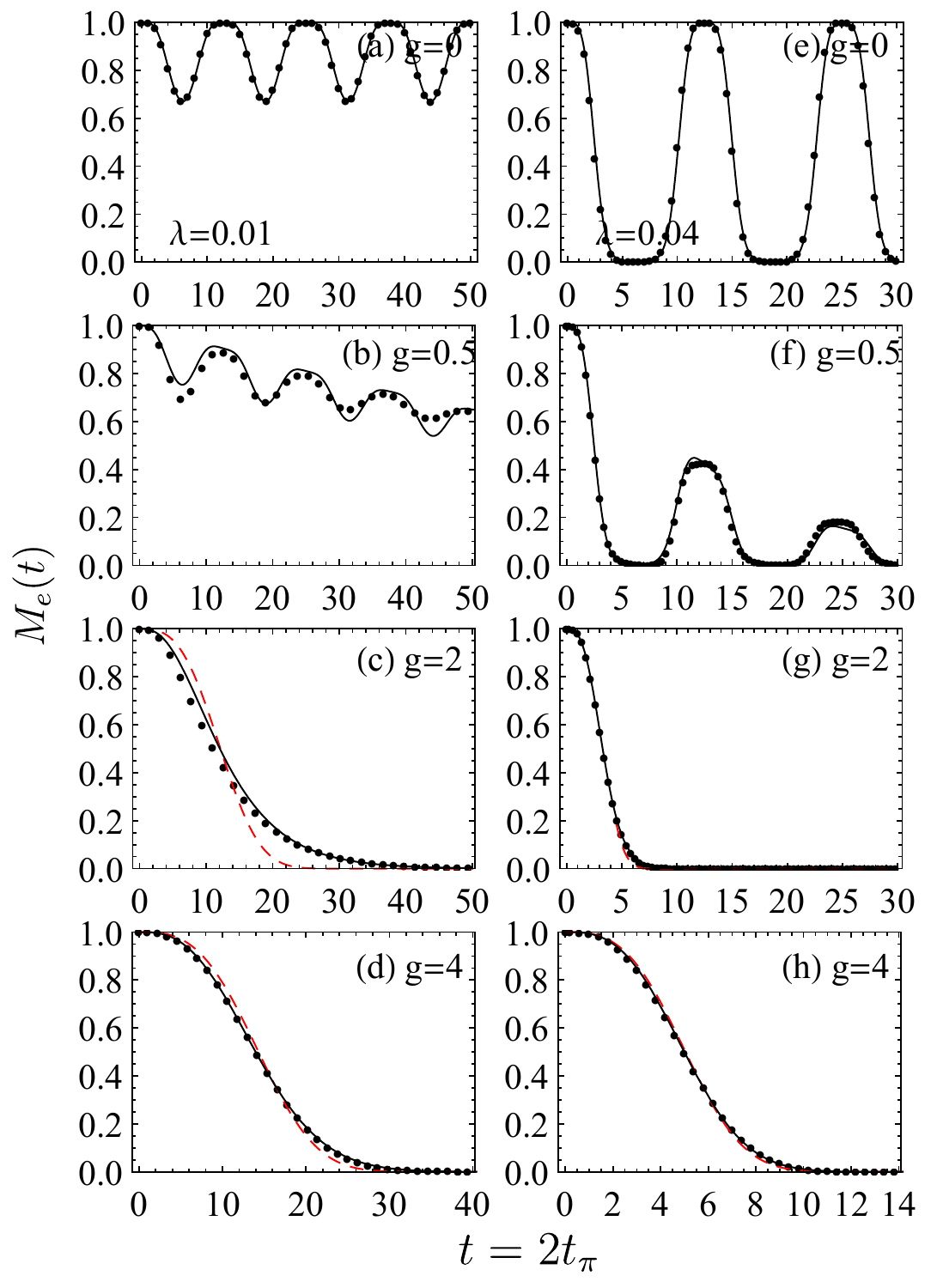}\\
  \caption{Numerical results of the transverse spin dynamics beyond relaxation time approximation for different interaction strength $g=\tilde{g}/(\hbar\tilde{\omega} l)$ (dotted). The black lines are the fitting to our ansatz Eq. (\ref{eq:ansatz}). The crossover from spin wave oscillation to diffusion dominated dynamics is observed in (b-c) and (f-g). The fully diffusion physics is revealed in (d) and (h). The dashed red lines are the fitting to $M_e^0(t)=\exp(-D\lambda^2t^3/12)$. The magnetic field gradient is chosen $\lambda=\tilde{\lambda}l/(\hbar\tilde{\omega})=0.01$ in (a-d) and $0.02$ in (e-h). The deviation to homogeneous solution is clearly shown in (c) and (d).}
  \label{fig:fig3}
\end{figure}

We show our numerical results for different interaction strength at magnetic field gradient $\lambda=0.01$ and $0.04$ in Fig.\ref{fig:fig3}. In non-interaction case $g=0$, the spin wave oscillations in Fig. \ref{fig:fig3}(a) and (e) agree well with our theoretical prediction of Eq. (\ref{eq:nonint}). Interactions cause the oscillation to decay, as the collisions damp out the spin waves. Fig. \ref{fig:fig3}(d) and (h) show that at strong interactions the behavior is purely diffusive. The curves in Fig. \ref{fig:fig3} are somewhat reminiscent of a damped harmonic oscillator, displaying undamped behavior in (a), (e), underdamped in (b), (f), and overdamped in (d), (h). In (c-d), (g-h), we also compare to the homogeneous solution $M_e^0(t)=\exp(-D\lambda^2t^3/12)$ (dashed red lines), treating $D$ as a free parameter chosen to best fit the numerical result. The fit is better for larger $g$ and smaller $\lambda$.
\begin{figure}[h]
  \centering
  \includegraphics[width=0.43\textwidth]{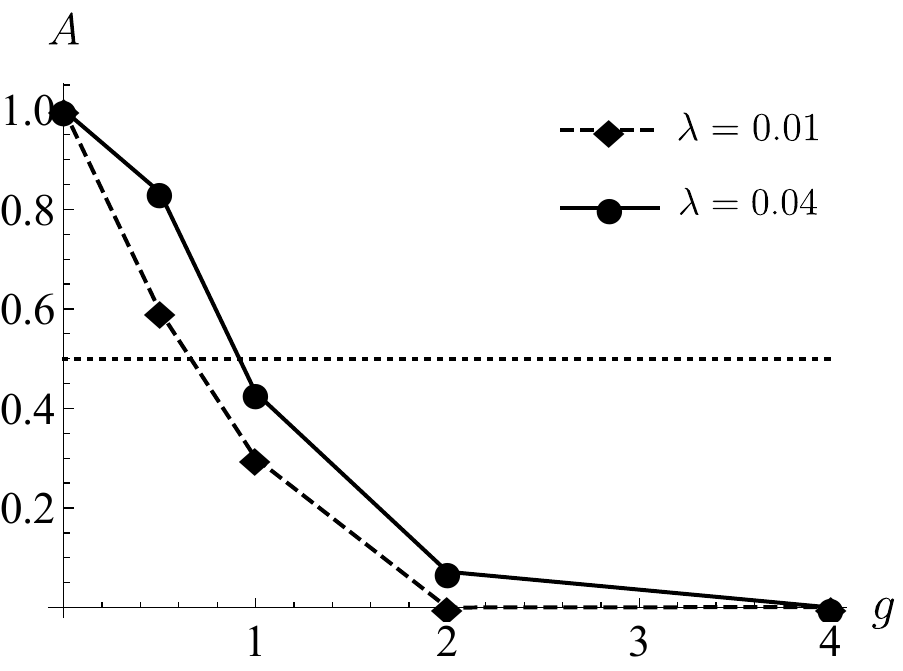}\\
  \caption{The contribution of spin wave oscillations as a function of interaction strength $g$ for magnetic field gradient $\lambda=\tilde{\lambda}l/(\hbar\tilde{\omega})=0.01$ (black squares) and $0.04$ (black circles). The lines are guides to the eyes. The dotted line labels the crossover.}
  \label{fig:fig4}
\end{figure}

To characterize the crossover from spin waves to spin diffusion, we use the following simple ansatz
\begin{align}
M_e(t=2t_\pi)=e^{-8A\lambda^2T\sin(t/4)^4-(1-A)Q\lambda^2},
\label{eq:ansatz}
\end{align}
where $A$ is a fit parameter and $Q$ is defined in our previous section. We show the best fit as solid lines in Fig. \ref{fig:fig3}. Despite its simplicity, this ansatz agrees well with our numerical results. When $A=1$, this ansatz is our collisionless solution, while when $A=0$ it is our hydrodynamic solution. To quantify the crossover, we plot the fit parameter $A$ in Fig. \ref{fig:fig4}. The crossover is relatively broad. There is a small dependence on $\lambda$, with $A$ being systematically larger for $\lambda=0.04$ than for $\lambda=0.01$.

\section{Conclusions}
To summarize, we have studied the transverse spin dynamics of polarized Fermi gases in a harmonic trap at a high temperature limit. We show how the spin dynamics evolve from collisionless to collisional. We give analytical expression for the magnetization. These analytic expressions agree well with our numerical solutions of the linearised quantum Boltzmann equation, in the relevant regimes.

Our approach, while powerful, is limited to describing the high temperature limit. At lower temperatures one would need to include the Hartree-Fock mean-field terms which drive the Leggett-Rice effect \cite{Thywissen2}, and demagnetization \cite{Koller2}. Our linearization of the collision integral will break down at large $\lambda$, where the dynamics create spin configuration which are far from equilibrium. Finally, to model lower temperatures, one may need to include ordering.

\section{Appendix: The spin wave dynamics in the non-interaction limit}
\label{app:collisionless}
The transverse distribution function for the non-interaction gas evolves as
\begin{align}
(\partial_t+v\partial_x-x\partial_v+i\lambda x)f_\perp(x,v,t)=0,
\label{eq:app1}
\end{align}
where we take the initial state to be $f_\perp(x,v,0)=e^{(\mu-x^2/2-v^2/2)/T}$. At time $t=t_\pi$, a $\pi$ pulse reverses the spins. Equivalently, we leave the spins unchanged but flip $\lambda$, taking $\lambda\to-\lambda$ for $t>t_\pi$. Eq. (\ref{eq:app1}) is satisfied by the ansatz
\begin{align}
f_\perp(x,v,t)=&\exp\left\lbrace\left[\mu-(x+iv_0(t))^2/2-(v+ix_0(t))^2/2\right.\right.\nonumber\\
&\left.\left.-x_0(t)^2/2-v_0(t)^2/2\right]/T\right\rbrace,
\end{align}
where
\begin{align}
&x_0(t)=\left\{\begin{array}{cc}
\lambda T(\cos t-1) & t\leq t_\pi \\
\lambda T\left[1+\cos t-2\cos(t-t_\pi)\right] &t> t_\pi
\end{array}\right.,\\
&v_0(t)=\left\{\begin{array}{cc}
\lambda T\sin t & t\leq t_\pi \\
\lambda T\left[\sin t-2\sin(t-t_\pi)\right] &t> t_\pi
\end{array}\right.
\end{align}
This structure is similar to the zero-temperature result in \cite{Xu}. The transverse magnetization is then measured at time $t=2t_\pi$. We find that in the final state $x_0^e(t=2t_\pi)=\lambda T\left[1+\cos t-2\cos (t/2)\right]$, $v_0^e(t=2t_\pi)=\lambda T\left[\sin t-2\sin (t/2)\right]$ and the transverse magnetization becomes
\begin{align}
M_e(t=2t_\pi)=\frac{1}{2\pi N}\int f_\perp(x,v,t) dxdv=e^{-8\lambda^2 T\sin(t/4)^4}.\nonumber
\end{align}

\section{Appendix: The solution of diffusion equation by the method of characteristics}
\label{app:method}
To solve the diffusion equation in Eq. (\ref{eq:trapdiff}), we first introduce a function $F_\lambda(x,t)$ with
\begin{align}
&m(x,t\leq t_\pi)=\int m(x_0,0)F_{\lambda}(x,t)dx_0,\nonumber\\
&m(x,t>t_\pi)=\int m(x_0,t_\pi)F_{-\lambda}(x,t-t_\pi)dx_0.\nonumber
\end{align}
From the above definition, it follows that $F_{\lambda}(x,0)=\delta(x-x_0)$. The diffusion equation can then be reduced to
\begin{align}
(\partial_t-D\partial^2_x+i\lambda x)F_\lambda-\tau\partial_x(xF_\lambda)=0,
\label{eq:diffF}
\end{align}
with the initial condition $F_\lambda(x,0)=\delta(x-x_0)$. We will analytically calculate $F_{\lambda}(x,t)$, then use it to produce the magnetization $m(x,t)$ from the initial magnetization $m(x,0)$.

To solve Eq. (\ref{eq:diffF}), we first transform the equation into Fourier space with $f_\lambda(k,t)=\int F_{\lambda}(x,t)e^{-ikx}dk$ as
\begin{align}
\frac{\partial f_\lambda}{\partial t}+(\tau k-\lambda)\frac{\partial f_\lambda}{\partial k}=-Dk^2f_\lambda.
\label{eq:momdiff}
\end{align}
The initial condition becomes $f_\lambda(k,0)=e^{-ikx_0}$.

This first-order partial differential equation and can be solved by the method of characteristics. The goal of this method is to find the characteristic curve $k(t)$ so that Eq. (\ref{eq:momdiff}) along this curve reduces to an ordinary differential equation (ODE). In particular, the characteristic curve obeys
\begin{align}
\frac{dk(t)}{dt}=\tau k(t)-\lambda,
\label{eq:curve}
\end{align}
giving an ODE
\begin{align}
\frac{df_\lambda(t)}{dt}\equiv\frac{\partial f_\lambda(t)}{\partial t}+\frac{\partial f_\lambda(t)}{\partial k(t)}\frac{dk(t)}{d t}=-Dk(t)^2f_\lambda(t).
\label{eq:ode}
\end{align}

Solving Eq. (\ref{eq:curve}) we find the characteristic curve
\begin{align}
k(t)=\frac{\lambda-e^{\tau t}\lambda+e^{\tau t}k_0\tau}{\tau},
\label{eq:k(t)}
\end{align}
where $k(0)=k_0$. The ODE (\ref{eq:ode}) with the initial condition $f_\lambda(0)=e^{-ik_0x_0}$ has solution
\begin{align}
f_{\lambda}(t)=\exp\left[-\frac{2ik_0x_0\tau^3+D(M_0+M_1\lambda+M_2\lambda^2)}{2\tau^3}\right],
\nonumber
\end{align}
where $M_0=(-1+e^{2\tau t})k_0^2\tau^2$, $M_1=-2(-1+e^{\tau t})^2k_0\tau$, $M_2=3-4e^{\tau t}+e^{2\tau t}+2\tau t$. The solution to Eq. (\ref{eq:momdiff}) is recovered by changing $k_0$ to $k(t)$ using Eq. (\ref{eq:k(t)}). Finally, we get the solution to Eq. (\ref{eq:diffF}) as
\begin{align}
F_\lambda(x,t)=\frac{1}{\sqrt{A}}e^{-B_0-B_1\lambda+B_2\lambda^2},
\label{eq:hight}
\end{align}
where $A=2\pi D/\tau(1-e^{-2\tau t})$, $B_0=\tau/(4D)(x_0-e^{t\tau}x)^2\left[\coth(\tau t)-1\right]$, $B_1=i/\tau(x+x_0)\tanh(\tau t/2)$, $B_2=D/\tau^3\left[2\tanh(\tau t/2)-\tau t\right]$.

\acknowledgements
We are grateful to Erich Mueller for many suggestions on our manuscript. We acknowledge the comments from Andrew Koller. This research is supported by the National Key Basic Research Program of China (No. 2013CB922002) and the National Natural Science Foundation of China (No. 11574028, 11504021). J.X. is also supported by CPSF (No. 2015M580043, 2016T90032), and FRFCU (No. FRF-TP-15-040A1).

\end{document}